\documentclass[aps,pra,superscriptaddress,amsmath,amssymb,preprintnumbers,floatfix,showpacs,12pt]{revtex4}
\usepackage{amssymb}
\usepackage{epsfig}

\begin{document}

\title{ Higher-order squeezing for the codirectional Kerr
nonlinear coupler}
\author{ Faisal A. A. El-Orany and J. Pe\v {r}ina }

\affiliation{Department of Optics, Palack\'{y} University, 17.
listopadu 50,
 772 07 Olomouc, Czech Republic}

\date{\today}

%\maketitle

\begin{abstract}
In this Letter we study the evolution of the higher-order
squeezing, namely, $n$th-order single-mode squeezing, sum- and
difference-squeezing for the codirectional Kerr nonlinear coupler.
We show that the amount of squeezing decreases when $n$, i.e. the
squeezing  order, increases. For specific values of the
interaction parameters squeezing factors exhibit a series of
revival-collapse phenomena, which become more pronounced when the
value of $n$ increases. Sum-squeezing can provide amounts of
squeezing greater than those produced by the $n$th higher-order
($n\geq 2$) squeezing for the same values of interaction
parameters and can map onto amplitude-squared squeezing. Further,
we prove that the difference-squeezing is not relevant measure for
obtaining information about squeezing from this device.

\end{abstract}
\maketitle {\bf Key words:} Squeezing;
Kerr nonlinear
 coupler.

\section{Introduction}
The optical coupler is a device composed of two (or more)
waveguides, which are placed close enough to each others. The
guided modes are coupled by means of the evanescent waves and
hence the energy exchanged between the waveguides can be
controlled \cite{jen1}. In this regard  directional coupler is an
important device for data transmission and optical communication
networks \cite{EKer1}. Such device has been experimentally
implemented, e.g. in \cite{exp1}. Directional coupler involving
Kerr nonlinearity is an important device owing to its application
in optics as an intensity-dependent routing switch
\cite{jen1,{qu10}}. The quantum properties for this device have
been studied by several authors
\cite{hora1,{hora2},{qu14},{qu15},{qu18},{faisal1},{ar1},{ar2}}.
For more details the reader can consult the review paper
\cite{qu20}.

As is well known that squeezed light has less noise than coherent
light in one of the field quadratures provided that the
uncertainty relation is fulfilled. This light has various
application, e.g. in quantum information, high precision
measurements, etc. This encourages researchers for developing
different types of squeezing. For instance, higher-order squeezing
of a single-mode case was suggested and examined in \cite{hong}.
In this direction the definitions for amplitude-squared squeezing
\cite{hill}, amplitude-cubed squeezing \cite{cub} and the $n$th
power squeezing \cite{zhi} have been developed. Furthermore,
higher-order two-mode squeezing has been given in the sense of the
sum- and difference-squeezing \cite{hill2}. Actually, the term
higher-order is given for the sum- and difference-squeezing since
the quadrature operators are defined in terms of a product (not a
sum) of mode operators. Quite recently, general multimode
sum-squeezing \cite{ngu1} and difference-squeezing \cite{ngu2}
have been adopted. Such definitions for  higher-order squeezing
are motivated by the development in the higher-order correlation
measurement techniques aiming to extract information efficiently
from the optical signal \cite{hong}. It is worth mentioning that
the first experimentally observed squeezed states are of the
two-mode type \cite{slu}. Also like one that exhibits second-order
(normal) squeezing, a field that is squeezed to a higher order is
 a pure quantum mechanical light and has no classical
description.

Generally, the earlier investigation given to CKNC  has been
entirely focused on the normal squeezing, e.g. \cite{qu14,qu20}.
 Thus in this
Letter we investigate the evolution of the higher-order squeezing
involving  also the sum- and difference-squeezing. This will be
done in the following order. In section 2 we give the basic
relations and equations, which will be used in the paper. In
section 3 we investigate and discuss the results. In section 4 we
give the main conclusions.

%%%%%%%%%%%%%%%%%%%%%%%%%%%%%%%%%%%%%%
\section{Basic equations and relations}
%%%%%%%%%%%%%%%%%%%%%%%%%%%%%%%%%%%%%%
In this section we give the basic equations and relations, which
include the Hamiltonian formalism for the system under
consideration, the solutions for the equations of motion and  the
general definition for squeezing.

 The Hamiltonian for the codirectional Kerr nonlinear
coupler (CKNC) is
\begin{equation}
\frac{\hat{H}}{\hbar }=\sum_{j=1}^{2}
[\omega _{j}\hat{a}_{j}^{\dagger }%
\hat{a}_{j}+\chi \hat{a}_{j}^{\dagger 2}
\hat{a}_{j}^{2}]+\widetilde{\chi }\hat{a}_{1}^{\dagger
}\hat{a}_{1}\hat{a}_{2}^{\dagger
}\hat{a}_{2}+\kappa (%
\hat{a}_{1}^{\dagger }\hat{a}_{2}+\hat{a}_{2}^{\dagger
}\hat{a}_{1}), \label{2}
\end{equation}
where $\omega_1$ and $\omega_2$ are the frequencies of the first
and the second modes  with the annihilation operators $\hat{a}_1$
and $\hat{a}_2$, respectively, $\chi$ and $\widetilde{\chi}$ are
the coupling constants proportional to the third-order
susceptibility $\chi^{(3)}$ and responsible for the self-action
and cross-action processes, respectively, $\kappa$ is the linear
coupling constant between the waveguides. The solution of the
Heisenberg equations related to (\ref{2}) when
$\widetilde{\chi}=2\chi$, can be easily obtained as:
\begin{eqnarray}
\begin{array}{lr} \hat{a}_{1}(t)=\exp
(-i\hat{\Lambda}t/2)\Bigl\{
\hat{%
a}_{1}(0)\left[ \cos (\lambda t)-i\frac{\Delta }{2\lambda }\sin
(\lambda t) \right] -i\frac{\kappa }{\lambda }\hat{a}_{2}(0)\sin
(\lambda t)\Bigr\},\\
 \\
\hat{a}_{2}(t) =\exp (-i\hat{\Lambda}t/2)\Bigl\{
\hat{%
a}_{2}(0)\left[ \cos (\lambda t)-i\frac{\Delta }{2\lambda }\sin
(\lambda t)\right] -i\frac{\kappa }{\lambda }\hat{a}_{1}(0)\sin
(\lambda t)\Bigr\},
  \label{4}
\end{array}
\end{eqnarray}
where $\lambda =\sqrt{\kappa ^{2}+\frac{1}{4}\Delta^{2}}$,
$\hat{\Lambda}=\omega_1+\omega _2+4\chi (\hat{a}_{1}^{\dagger
}\hat{a}_{1}+\hat{a}_{2}^{\dagger } \hat{a}_{2})$ and $\Delta$ is
the frequency mismatch. It is obvious that
$\hat{a}_{1}(t)\leftrightarrow \hat{a}_{2}(t)$ when
$\hat{a}_{1}(0)\leftrightarrow \hat{a}_{2}(0)$. The nature of the
coupler, i.e. the switching of energy between waveguides,
manifests itself as periodic functions in (\ref{4}). Moreover, in
addition with the energy exchange, both optical fields in the CKNC
undergo the self-phase modulation owing to nonlinearity in the
waveguides described by the cubic susceptibility $\chi^{(3)}$ and
would manifest itself in the equations as a nonlinear-modulation
phase term, as we shall see. Assuming that the two modes are
initially prepared in the coherent light
$|\alpha_1,\alpha_2\rangle$,
 one can evaluate the general form for
the different moments of the operators $\hat{A}_{j}(t)=
\hat{a}_{j}(t)\exp[\frac{it}{2}(\omega_1+\omega_2)]$, where
$\hat{a}_{j}(t)$ are given by (\ref{4}), as
\begin{eqnarray}
\begin{array}{lr}
\langle \hat{A}_{1}^{\dagger n_1}(t) \hat{A}_{2}^{\dagger n_3}(t)
\hat{A}_{1}^{n_2}(t) \hat{A}_{2}^{n_4}(t)\rangle=
\exp[(|\alpha_{1}|^{2}
+|\alpha_{2}|^{2})(z^{n_2+n_4-n_3-n_1}-1)]\\
\\
\times \bar{\alpha}_1^{n_2}(t) \bar{\alpha}_2^{n_4}(t)
\bar{\alpha}_1^{* n_1}(t) \bar{\alpha}_2^{* n_3}(t)
z^{[n_2n_4+\frac{n_2}{2}(n_2-1)+\frac{n_4}{2}(n_4-1)
-n_1n_3-\frac{n_1}{2}(n_1-1)-\frac{n_3}{2}(n_3-1)]},\label{f1}
\end{array}
\end{eqnarray}
where $n_j,j=1,2,3,4$ are integers,
\begin{equation}
z=\exp(-2i\chi t), \quad \bar{\alpha}_{1}(t)=
\bar{\alpha}_{x}(t)+i\bar{\alpha}_{y}(t), \quad
\bar{\alpha}_{2}(t)= \bar{\alpha}'_{x}(t)+i\bar{\alpha}'_{y}(t)
 \label{f2}
\end{equation}
and
\begin{eqnarray}
\begin{array}{lr}
\bar{\alpha}_{x}(t)= \alpha_{1}\cos \lambda t, \quad
\bar{\alpha}_{y}(t)= -[\alpha_{1}\frac{\Delta }{2}
+\alpha_2\kappa ]\frac{\sin \lambda t}{\lambda },\\
\\
\bar{\alpha}'_{x}(t)= \alpha_{2}\cos \lambda t, \quad
\bar{\alpha}'_{y}(t)= -[\alpha_{2}\frac{\Delta }{2}
+\alpha_1\kappa]\frac{\sin \lambda t}{\lambda }.
 \label{ff2}
\end{array}
\end{eqnarray}
We have assumed that $\alpha_1$ and $\alpha_2$ are real.

On the other hand, for investigating squeezing we have to define
two quadratures $\hat{X}$ and $\hat{Y}$, which denote the real
(electric) and imaginary (magnetic) parts of the radiation field.
Assume that these quadratures satisfy the following commutation
rule:
\begin{equation}
[\hat{X},\hat{Y}]=\frac{\hat{C}}{2}, \label{ol1}
\end{equation}
where $\hat{C}$ may be $c$-number or operator. The uncertainty
relation associated with the commutation rule (\ref{ol1}) is
\begin{equation}
\langle (\triangle \hat{X})^{2}\rangle \langle (\triangle \hat{Y}
)^{2}\rangle \geq \frac{|\langle \hat{C}\rangle|^{2}}{16},
\label{ol2}
\end{equation}
where  $\langle (\triangle \hat{X})^{2}\rangle=\langle
\hat{X}^{2}\rangle -\langle \hat{X}\rangle ^{2}$ and similar form
can be given for $\langle (\triangle \hat{Y})^{2}\rangle$. The
system is said to be squeezed in the $X$-quadrature if

\begin{equation}
 S=\frac{4\langle (\triangle \hat{X}(t))^2\rangle -
|\langle \hat{C}\rangle|}{|\langle \hat{C}\rangle|}\leq 0.
\label{ol3}
\end{equation}
The equality sign in (\ref{ol3}) holds  for minimum-uncertainty
states.
 Similar definition can be given for the $Y$-quadrature
(defining a $Q$-factor). Equations (\ref{f1})--(\ref{ol3}) provide
all the necessary tools to describe evolution of the different
types
 of higher-order squeezing.

%%%%%%%%%%%%%%%%%%%%%%%%%%%%%%%%%
\section{Results and discussions}
%%%%%%%%%%%%%%%%%%%%%%%%%%%%%%%

In this section we discuss three types of squeezing, which are
$n$th-order single-mode squeezing, sum-squeezing and
difference-squeezing. This will be performed in the following
parts.
%%%%%%%%%%%%%%%%%%%%%%%%%%%%%%%%%%%%%%%%%%%%%%%%
 \subsection{The $n$th-order single-mode squeezing}
 %%%%%%%%%%%%%%%%%%%%%%%%%%%%%%%%%%%%%%%%%%%%%%%%%
In this part we treat the $n$th-order single-mode squeezing. For
convenience we use the definition given in  \cite{zhi}.
  In this
 case $\hat{X}, \hat{Y}$ and $\hat{C}$ take the forms:

\begin{equation}
 \hat{X}=\frac{1}{2}[\hat{A}^{n}_{1}(t)+
\hat{A}^{\dagger n}_{1}(t)], \quad
 \hat{Y}=\frac{1}{2i}[\hat{A}^{n}_{1}(t)-
\hat{A}^{\dagger n}_{1}(t)],\quad \hat{C}=
\hat{A}^{n}_{1}(t)\hat{A}^{\dagger n}_{1}(t)- \hat{A}^{\dagger
n}_{1}(t)\hat{A}_{1}^{n}(t),  \label{ol5}
\end{equation}
where $n$ is positive integer. For obtaining some accurate
information we assume that $\alpha_1=\alpha_2=\alpha$ and
$\Delta=0$ (resonance case). From (\ref{f1}), (\ref{ol3}) and
(\ref{ol5}) we can obtain
\begin{equation}
 S_1(t)=\mu [1+h_1(t) -h_2(t)],\qquad
 Q_1(t)=\mu [1-h_1(t) -h_3(t)],
 \label{ol7}
\end{equation}
where
\begin{eqnarray}
\begin{array}{lr}
h_1(t)= \cos [2\lambda t n+2n(2n-1)\chi t+\epsilon\sin(4n\chi t)]f(2n\chi t),\\
\\
h_2(t)=2\cos^2 [\lambda t n+n(n-1)\chi
t+\epsilon\sin(2n\chi t)]f^2(n\chi t),\\
\\
h_3(t)=2\sin^2 [\lambda t n+n(n-1)\chi t+\epsilon\sin(2n\chi
t)]f^2(n\chi t)
 ,\\
\\
\epsilon=|\alpha_1|^2+|\alpha_2|^2,\qquad \mu=\frac{2\alpha^{2n}}{\langle\hat{C}\rangle},\\
\\
f(n\chi t)=\exp[-2\epsilon\sin^2(n\chi t)].
 \label{ol8}
 \end{array}
\end{eqnarray}
Actually, the origin of occurrence of the nonclassical effects in
CKNC is in the existence of the  envelope function and/or of the
nonlinear-modulation phase term $f(n\chi t)$. The evolution of
this function is  mainly responsible for the features of the
squeezing factors. In this regard the value of the parameter $\chi
t$ plays the crucial role in obtaining squeezing. The envelope
function is periodic and the period decreases as the value of $n$
 increases. Roughly speaking, it is obvious that squeezing occurs
when the amount in the brackets of (\ref{ol7})--which is
finite--is less than zero. The pre-factor $\mu$ plays an
amplification role. Suppose that $ \chi t=\frac{\pi}{m}$. Thus for
$m=n$ or $m=1$ the system reduces to its initial stage, i.e.
disentangled coherent states but the amplitudes may be different
from those of the initial ones. Consequently, squeezing may occur
and switching between the two waveguides only when $n/m=l, l$ is
fraction. Now we prove that for specific values of interaction
time the system can exhibit higher-order squeezing. For instance,
for $n$  odd, i.e. $n(2n-1)$ is odd and $n(n-1)$ is even, and
$\chi t=\pi/2$, say, the expressions (\ref{ol7}) reduces to
%%%%%%%%%%%%%%%%%%%%%%%%%%%%%%%%%%%%%%%%%%%%%%%%%%%%%%%%%%%%%%%
\begin{figure}
 \includegraphics[width=1.0\linewidth]{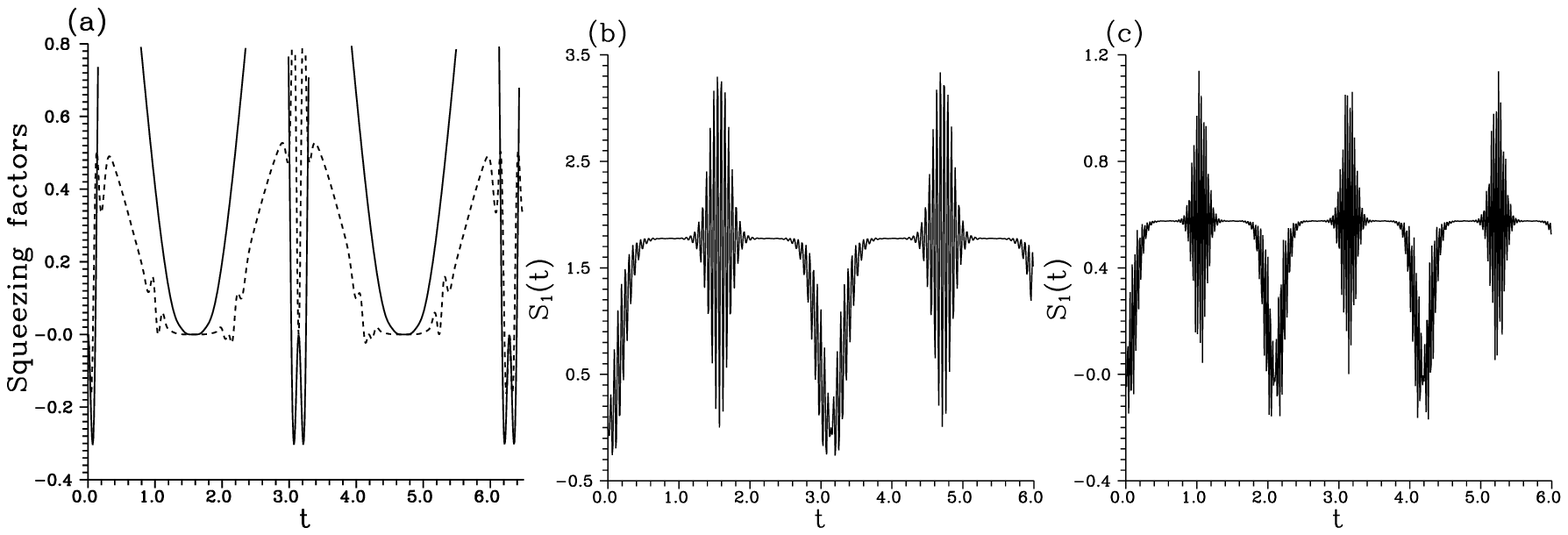}
 \caption{
Evolution of the squeezing factor $S_1(t)$  of the first mode when
 $\kappa=1, \chi =0.5, (\alpha_1,\alpha_2)=(2,0)$ and for
(a) $\Delta=0$ (solid curve for $n=2$ and dashed curve for $n=3$),
(b) $(\Delta,n)=(50,2)$, and (c) $(\Delta,n)=(50,3)$.}
\end{figure}
%%%%%%%%%%%%%%%%%%%%%%%%%%%%%%%%%%%%%%%%%%%%%%%%%%%%%%%%%%%%

\begin{eqnarray}
\begin{array}{lr}
 S_1(t)=\mu\left\{1-\cos (2n\lambda t)-2\cos^2(n\lambda t)
 \exp(-2\epsilon)
\right\}, \\
 \\
Q_1(t)=\mu\left\{1+\cos (2n\lambda t)-2\sin^2(n\lambda t)
\exp(-2\epsilon) \right\}.
 \label{ol14}
 \end{array}
\end{eqnarray}
Therefore, for $\lambda t=\pi/n$, say, squeezing can be only
observed in $S_1(t)$, whereas for $\lambda t=\pi/(2n)$ it is only
obtained in $Q_1(t)$. One can realize that the values of
interaction time to which squeezing occur, depend on the order of
the squeezing. On the other hand, when $n$ is even and $\chi
t=\pi/2$ one can prove that the system provides its initial stage,
i.e. disentangled coherent states. As we mentioned above the
quantity in the brackets of (\ref{ol7}) is finite, i.e. its value
locates in the intervals $[0,\pm 2]$, so the natural question is
that which value of $n$ provides maximum squeezing? The answer for
this can be obtained by examining the amplification factor $\mu$
for different values of $n$. We found that as $n$ increases the
value of $\mu$ decreases. This means that the best value of
squeezing can be obtained for the lowest order, i.e. $n=1$ for
normal squeezing \cite{qu14}. Nevertheless, the periodicity of
occurring squeezing in the time domain increases as $n$ increases.

%%%%%%%%%%%%%%%%%%%%%%%%%%%%%%%%%%%%%%%%%%%%%%%%%%%%%%%%%%%%%%%
\begin{figure} \includegraphics[width=1.0\linewidth]{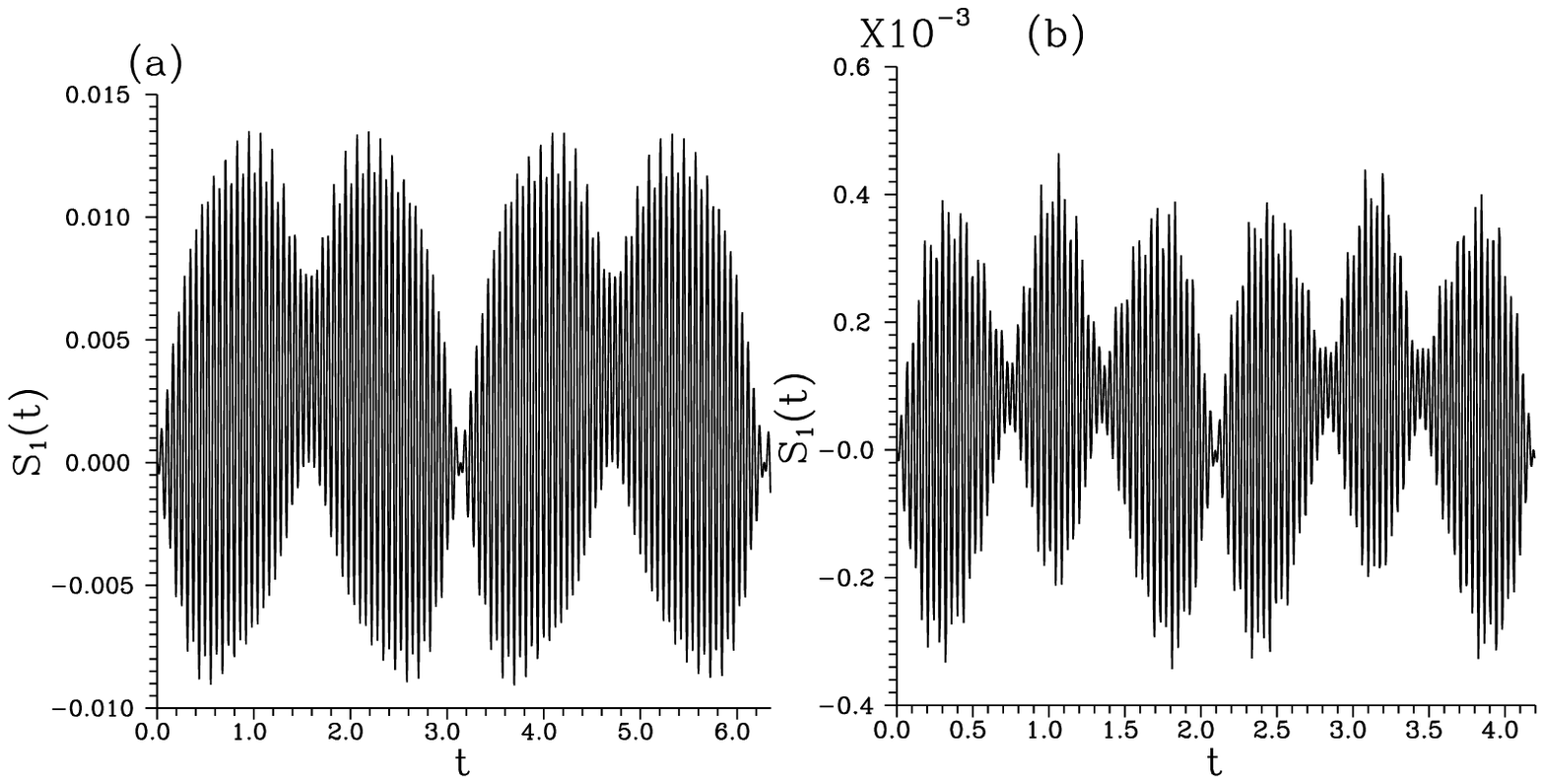}
\caption{
Evolution of the squeezing factor $S_1(t)$  of the first mode when
 $\kappa=1, \chi =0.5, (\alpha_1,\alpha_2)=(0.3,0.3), \Delta=50$, and for
 (a) $n=2$, (b) $n=3$.}\end{figure}
%%%%%%%%%%%%%%%%%%%%%%%%%%%%%%%%%%%%%%%%%%%%%%%%%%%%%%%%%%%%

These facts are remarkable in Figs. (1) and (2) for strong and
weak intensities, respectively, for given values of interaction
parameters. From Fig. 1(a) one can observe that squeezing occurs
periodically and the periodicity increases as well as the amount
of squeezing decreases when the order $n$ increases (compare the
dashed and solid curves). The value of the detuning parameter
$\Delta$ plays an important role (see Figs. 1(b) and (c)). From
Figs. 1(b) and (c) for $\Delta >>1$, $S_1(t)$ exhibits particular
shape of periodic revival-collapse phenomenon and the amount of
the nonclassical squeezing becomes much more pronounced than
before (compare Fig. 1(b) and (c) to the solid and dashed curves
in Fig. 1(a), respectively). Generally, the occurrence of
revival-collapse phenomenon in the evolution of squeezing factors
may be explained as follows. Basically $S_1(t)$ includes two forms
of periodic function, namely, the trigonometric and envelope
functions. These functions are periodic in the two parameters
$\lambda t$ (with period $\pi/n\lambda$) and $\chi t$ (with period
$\pi/n\chi$). Generally, when the values of $\Delta$ increase, the
period of the energy exchange between waveguides decreases, i.e.
many oscillations occur, till the interaction time becomes $t\chi
=\pi/2n$, at this moment the field is trapped instantaneously by
nonlinearity in the waveguides and the squeezing factors show
collapse. As the interaction proceeds the phenomenon is
periodically repeated. The sensitivity of the revival-collapse
phenomenon to the value of $n$ can be realized by comparing Fig.
1(b) and Fig. 1(c).  The number of the revival patterns increases
as $n$ increases.
 The particular shape of revival-collapse phenomenon in Figs.
1(b) and (c) can be understood as follows. In (\ref{ol7}) we have
two forms of the envelope function, which are $f(2n\chi t)$ and
$f(n\chi t)$. As we mentioned above these are periodic functions,
however, the period of the first one is two times less than that
of the second. Consequently, for $\chi t=\frac{m\pi}{n}, m$ is
integer,
 $h_1(t)$ and $h_2(t)$ provide simultaneously their maximum
 contribution, which interfere with each others producing
 squeezing in $S_1(t)$, whereas for
$\chi t=\frac{m\pi}{n}, m$ is odd integer, $h_1(t)$ provides
 the main contribution and hence complete
revivals occur. Such behavior can be realized  only for large
intensities, where $f(n\chi t)=0$ or $1$, however, for weak
intensities the behavior will be rather different. This is related
to the fact that $f(n\chi t)\simeq 1$ everywhere (compare Figs.
(1) and (2)). Now we draw the attention to Figs. 2(a) and (b).
Comparison between  Figs. 1 and 2 leads to the result that the
value of squeezing for strong-intensity regime  is much greater
than that for the weak-intensity regime and the shapes of
revival-collapse phenomenon in the two regimes are quite
different. The origin of this difference is that for weak
intensities there is no exact collapse causing that the revival
patterns are much broader than that of the strong-intensity
regime. Also the various facts mentioned above related to the
value of $n$ are still valid in the weak-intensity regime.
Throughout the discussion we have focused on the evolution of
$S_1(t)$ because we have noted that for $\Delta=0$, $Q_1(t)$ is
almost positive, i.e. it cannot provide squeezing, and for
$\Delta\neq 0$ it provides typical forms as those for the $S_1(t)$
(see Figs. 1(b), (c) and Figs. 2).

%%%%%%%%%%%%%%%%%%%%%%%%%%%%%%%%%%%%%%%%%%%%%%%%%%
\subsection{Sum-squeezing}
%%%%%%%%%%%%%%%%%%%%%%%%%%%%%%%%%%%%%%%%%%%%%%%%%
It is worth reminding that the sum- and difference-squeezing has
been realized in nonlinear optics for four-wave sum \cite{sum} and
difference \cite{dif} frequency generation. We proceed by using
the definition given in \cite{hill2} for sum-squeezing.
 In this case the operators
$\hat{X}, \hat{Y}$ and $\hat{C}$ take the forms :

\begin{equation}
 \hat{X}=\frac{1}{2}[\hat{A}_{1}(t)\hat{A}_{2}(t)+
\hat{A}^{\dagger}_{1}(t)\hat{A}^{\dagger}_{2}(t)],\quad
 \hat{Y}=\frac{1}{2i}[\hat{A}_{1}(t)\hat{A}_{2}(t)-
\hat{A}^{\dagger}_{1}(t)\hat{A}^{\dagger}_{2}(t)],\quad
\hat{C}=\hat{N}_1+\hat{N}_2+1.
 \label{sqz1}
\end{equation}
One can easily check that when $\alpha_1=\alpha_2$ sum-squeezing
map onto the amplitude-squared squeezing given above, i.e. $n=2$
(see Figs. 1). Here we pay attention to the case
$\alpha_1\neq\alpha_2$. Assume that $\alpha_1=\alpha, \alpha_2=0$
and $\triangle=0$. For this case the sum-squeezing factors can be
evaluated as
\begin{eqnarray}
\begin{array}{lr}
 S_2(t)=\frac{2\alpha^4}{\alpha^2+1} \sin^{2}(\lambda t)\cos^{2}(\lambda t)
 \Bigl\{1+
 \cos(12\chi t+\epsilon\sin(8\chi t))\exp[-2\epsilon\sin^{2}(
4\chi t)]\\
\\
-2 \sin^{2}(2\chi t+\epsilon\sin(4\chi t))\exp[-4\epsilon\sin^{2}(
2\chi t)]
\Bigr\},\\
 \\
Q_2(t)=\frac{2\alpha^4}{\alpha^2+1} \sin^{2}(\lambda
t)\cos^{2}(\lambda t)
 \Bigl\{1-
 \cos(12\chi t+\epsilon\sin(8\chi t))\exp[-2\epsilon\sin^{2}(
4\chi t)]\\
\\
-2 \cos^{2}(2\chi t+\epsilon\sin(4\chi t))\exp[-4\epsilon\sin^{2}(
2\chi t)] \Bigr\}.
 \label{ol10}
 \end{array}
\end{eqnarray}
Expressions (\ref{ol10}) show that the system is able to produce
sum-squeezing, e.g. when $\chi t=\pi/4$ those expressions reduce
to
\begin{eqnarray}
\begin{array}{lr}
 S_2(t)=-\frac{\alpha^4}{\alpha^2+1} \sin^{2}(2\lambda
 t)\exp(-2\epsilon),
 \\
Q_2(t)=\frac{\alpha^4}{\alpha^2+1} \sin^{2}(2\lambda
 t).
 \label{ol11}
 \end{array}
\end{eqnarray}
It is evident that squeezing can occur in the first quadrature.
Figs. 3(a) and (b) are given for $S_2(t)$ for strong and weak
intensity regimes, respectively, for the given values of the
parameters. One can observe that  squeezing occurs periodically
and becomes more pronounced when the intensities increase (see the
inset in Fig. 3(b), also compare Fig. 3(a) and Fig. 3(b)).
Influence of the detuning parameter is shown in Fig. 3(b), which
manifests itself as  the revival-collapse phenomenon. Explanation
as that given in the first part can be given here. Further, we
have noted that the nonclassical values of the sum-squeezing are
much greater than those for the $n$th-order single-mode squeezing
($n \geq 2$) for the same values of the interaction parameters.
Also for $\Delta>>1, \alpha_j>1$  Figs. 1(b) and (c) are obtained.
Finally, conclusions similar to those given for $Q_1(t)$ in the
first part are valid  for $Q_2(t)$.
%%%%%%%%%%%%%%%%%%%%%%%%%%%%%%%%%%%%%%%%%%%%%%%%%%
\subsection{Difference-squeezing}
%%%%%%%%%%%%%%%%%%%%%%%%%%%%%%%%%%%%%%%%%%%%%%%%%
In this part we show that difference-squeezing factors fail to
give information about squeezing from the coupler. For
difference-squeezing the operators $\hat{X}, \hat{Y}$ and
$\hat{C}$ take the forms \cite{hill2}:

\begin{equation}
 \hat{X}=\frac{1}{2}[\hat{A}_{1}(t)\hat{A}^{\dagger}_{2}(t)+
\hat{A}^{\dagger}_{1}(t)\hat{A}(t)],\quad
 \hat{Y}=\frac{1}{2i}[\hat{A}_{1}(t)\hat{A}^{\dagger}_{2}(t)-
\hat{A}^{\dagger}_{1}(t)\hat{A}(t)],\quad
\hat{C}=\hat{N}_2-\hat{N}_1.
 \label{dqz1}
\end{equation}
From (\ref{f1}), (\ref{ol3}) and (\ref{dqz1}) we can obtain the
difference-squeezing factors  as
\begin{eqnarray}
\begin{array}{lr}
 S_3(t)=2{\rm Re}[\bar{\alpha}^{2}_1(t)
\bar{\alpha}^{*2}_2(t)]+2|\bar{\alpha}_1(t)|^2
|\bar{\alpha}_2(t)|^2+2|\bar{\alpha}_1(t)|^2- \left[
\bar{\alpha}_x(t)\bar{\alpha}'_x(t)+
\bar{\alpha}_y(t)\bar{\alpha}'_y(t)\right]^2,\\
\\
Q_3(t)=-2{\rm Re}[\bar{\alpha}^{2}_1(t)
\bar{\alpha}^{*2}_2(t)]+2|\bar{\alpha}_1(t)|^2
|\bar{\alpha}_2(t)|^2+2|\bar{\alpha}_1(t)|^2- \left[
\bar{\alpha}_x(t)\bar{\alpha}'_y(t)-
\bar{\alpha}_x(t)\bar{\alpha}'_y(t)\right]^2,
 \label{ol13}
 \end{array}
\end{eqnarray}
where $\bar{\alpha}_j(t)$ are given by (\ref{f2}) and (\ref{ff2})
and ${\rm Re}$ stands for real value. It is evident that
(\ref{ol13}) is independent of the nonlinear-modulation phase term
and consequently the system cannot provide difference-squeezing.
This can be confirmed after minor manipulation with (\ref{ol13}),
which reduces to
\begin{equation}
S_3(t)=Q_3(t)=2|\bar{\alpha}_1(t)|^2. \label{ol14}
\end{equation}
This means that the two squeezing factors are typical and equals
twice the mean photon number in the first waveguide. Such behavior
of difference-squeezing can be understood by noting that the
coupler and the quadratures of the difference-squeezing  are
describing by the same mechanism. To be more specific, the
quadratures of the difference-squeezing represent  up conversion
processes (cf. \ref{dqz1}), i.e. when one photon is created in
the first mode the other is annihilated in the second mode. On the
other hand, the coupler is basically operating by switching energy
between waveguides (conservation of energy).

%%%%%%%%%%%%%%%%%%%%%%%%%%%%%%%%%%%%%%%%%%%%%%%%%%%%%%%%%%%%%%%
\begin{figure}
 \includegraphics[width=1.0\linewidth]{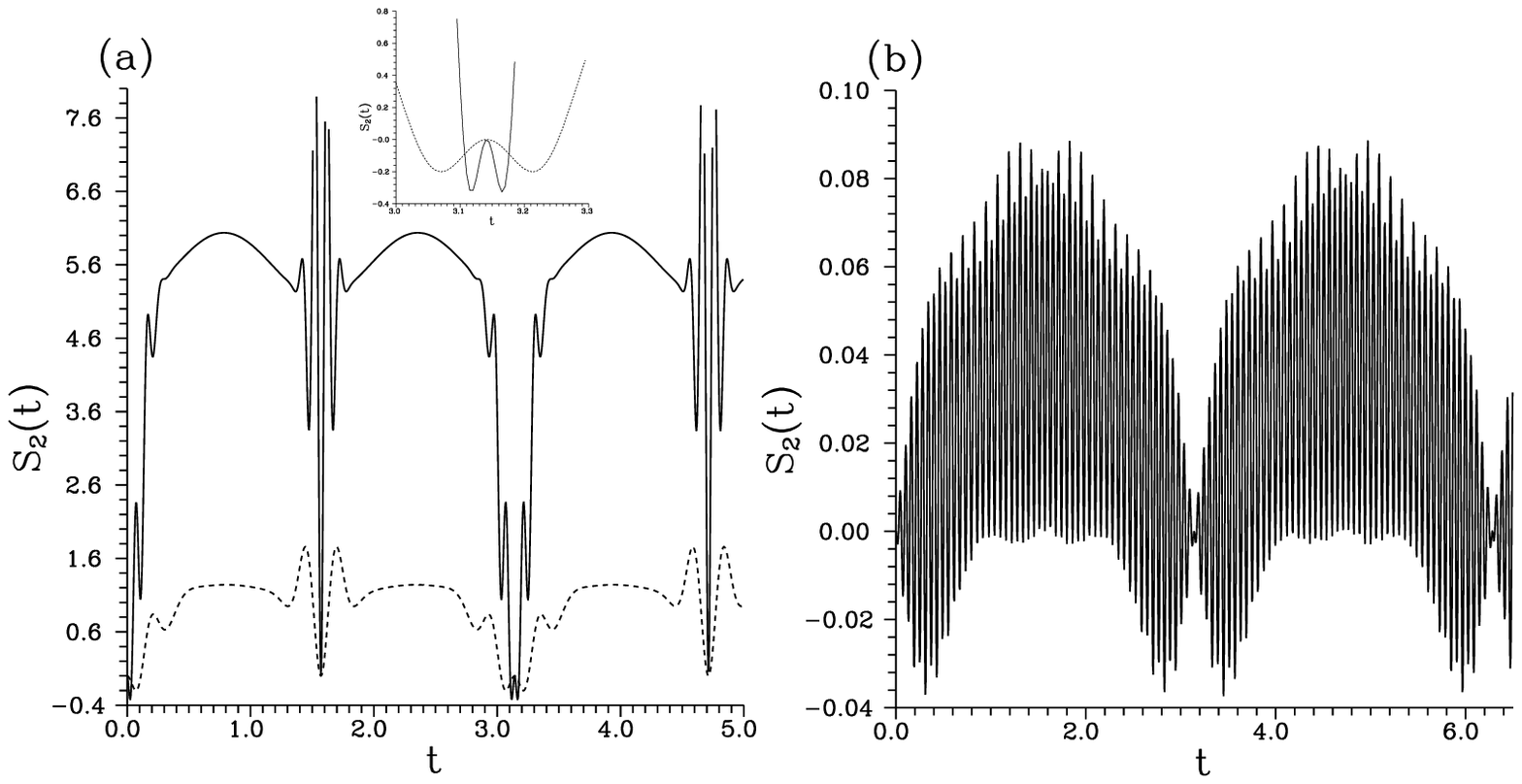}  \caption{
Evolution of the sum-squeezing factor $S_2(t)$ for
 $\kappa=1, \chi =0.5$, and for (a) $ (\Delta,\alpha_1,\alpha_2)=(0,1,1.5)$ (short-dashed curve) and $(0,2,3)$
 (solid curve), (b)
 $ (\Delta,\alpha_1,\alpha_2)=(50,0.3,0.6)$. The inset in (a) is given for the sake of comparison.}
\end{figure}
%%%%%%%%%%%%%%%%%%%%%%%%%%%%%%%%%%%%%%%%%%%%%%%%%%%%%%%%%%%%

\section{Conclusions}
%%%%%%%%%%%%%%%%%%%%%%%%%%
Throughout this Letter we have studied for the first time the
higher-order squeezing for CKNC. For the $n$th-order single-mode
squeezing we have found that the amount of squeezing decreases as
the order of the squeezing increases regardless of the values of
the intensities. Frequency mismatch can increase (or generate)
squeezing in the quadratures. Also squeezing factors exhibit
revival-collapse phenomenon resulting from the competition between
the Kerr nonlinearity and the frequency mismatch.
 Further,  the number of the revival patterns increase when the order
of squeezing increases. The locations of these revival patterns in
the time domain depend on the values of $nt\chi$, whereas their
shapes depend on the intensities of the field launched in the
waveguides initially. Furthermore, sum-squeezing can map onto
amplitude-squared squeezing when the intensities are equal and can
provide revival-collapse phenomenon based on the values of
$\Delta$. Sum-squeezing can provide amounts of squeezing greater
than those produced by the $n$th higher-order ($n\geq 2$)
squeezing for the same values of interaction parameters. This
means that the sum-squeezing is a better measure for extracting
information about squeezing from CKNC. These conclusions are in
relation to the structure of the nonlinear part of the Hamiltonian
 (\ref{2}). Also we have proved that the difference-squeezing is
not suitable for extracting information about squeezing from CKNC.
The final remark, we have numerically noted that the occurrence of
the revival-collapse phenomenon in the squeezing factors depends
on the value of $\Delta$ and not on $\kappa$, i.e. on the
intensity of the linear exchange between waveguides.

%%%%%%%%%%%%%%%%%%%%%%%%%%%%%%%%%%%%%%%%%%%%%%%%%%%%%%%%%%%%%
\section*{ Acknowledgement}
%%%%%%%%%%%%%%%%%%%%%%%%%%%%%%%%%%%%%%%%%%%%%%%%%%%%%%%%%%%%%%
 The authors  thank the partial support from the grant
LN00A015 of the Czech Ministry of Education and from the EU
Project COST OCP 11.003.

%%%%%%%%%%%%%%%%%%%%%%%%%%%%%%%%%%%%%%%%%%%%%%%%%%%%%%%%%%%%%%%%%%%%%%%%%%%%%%%%
%%%%%%%%%%%%%%%%%%%%%%%%%%%%%%%%%%%%%%%%%%%%%%%%%%%%%%%%%%%%%%%%%%%%%%%%%%%%%%%%
\section*{References}
%%%%%%%%%%%%%%%%%%%%%%%%%%%%%%%%%%%%%%%%%%%%%%%%%%%%%%%%%%%%%%%%%%%%%%%%%%%%%%%%
%%%%%%%%%%%%%%%%%%%%%%%%%%%%%%%%%%%%%%%%%%%%%%%%%%%%%%%%%%%%%%%%%%%%%%%%%%%%%%%%

\end{document}